\documentclass[11pt]{article}

\usepackage{amsmath,amssymb,color}
\usepackage{pstcol}
\usepackage{multicol}
\usepackage{graphicx}
\usepackage{amsfonts}
\newcommand{\rem}[1]{}
\newcommand{\remfigure}[1]{#1}
\newcommand{\bfi}{\bfseries\itshape}
\newcommand{\comment}[1]{\vspace{1 mm}\par 
\marginpar{\large\underline{}}\noindent
\framebox{\begin{minipage}[c]{0.98 \textwidth}
{\bfi\color{blue} #1} \end{minipage}}\vspace{1 mm}\par}

\DeclareMathAlphabet{\mathbi}{OML}{cmm}{b}{it} 
\newcommand{\bx}{\mathbi{x}}

\newcommand{\bel}{\begin{equation}\label}
\newcommand{\ee}{\end{equation}}
\newcommand{\ben}{\begin{enumerate}}
\newcommand{\een}{\end{enumerate}}
\newcommand{\bit}{\begin{itemize}}
\newcommand{\eit}{\end{itemize}}
\newcommand{\bB}{\mathbi{B}}
\newcommand{\bhB}{\boldsymbol{\hat{B}}}
\newcommand{\bF}{\mathbi{F}}
\newcommand{\bdb}{\mathbi{b}}

\newcommand{\ba}{\mathbi{a}}

\newcommand{\bD}{\mbox{\boldmath$\mathcal{D}$}}
\newcommand{\curr}{\mathbi{J}}

\newcommand{\bq}{\mathbi{q}}

\newcommand{\bfq}{\mathfrak{q}}

\newcommand{\bfw}{\mathfrak{w}}

\newcommand{\cast}{\circledast}

\newcommand{\bu}{\mathbi{u}}

\newcommand{\bw}{\mathbi{w}}
\newcommand{\bhw}{\mathbi{\hat{w}}}
\newcommand{\bom}{\mbox{\boldmath$\omega$}}

\newcommand{\bchi}{\mbox{\boldmath$\chi$}}
\newcommand{\bhchi}{\boldsymbol{\hat{\chi}}}

\newcommand{\bgamma}{\mbox{\boldmath$\gamma$}}

\newcommand{\beq}{\begin{eqnarray}\label} 
\newcommand{\eeq}{\end{eqnarray}} 
\newcommand{\bc}{\begin{center}} 
\newcommand{\ec}{\end{center}} 
\newcommand\shalf{\ensuremath{{\scriptstyle\frac{1}{2}}}}

\newtheorem{theorem}{Theorem}

\makeatletter
\@addtoreset{equation}{section}

\makeatother

\definecolor{mush}{cmyk}{.3,.1,.6,.1}
\color{black} 

\begin{document}

\title{Lagrangian analysis of alignment dynamics for isentropic compressible
magnetohydrodynamics}

\author{J. D. Gibbon$^1$ and D. D. Holm$^{1,2}$
\\$^1$
Department of Mathematics, Imperial College London SW7 2AZ, UK
\\{\footnotesize email: j.d.gibbon@ic.ac.uk and d.holm@ic.ac.uk}
\\
$^2$Computer and Computational Science,\\
Los Alamos National Laboratory,\\
MS D413 Los Alamos, NM 87545, USA
\\
{\footnotesize email: dholm@lanl.gov}
}

\date{6 August 2006}

\maketitle

\begin{abstract} 
After a review of the isentropic compressible magnetohydrodynamics (ICMHD) equations, a
quaternionic framework for studying the alignment dynamics of a general fluid flow is
explained and applied to the ICMHD equations.
\end{abstract}

\tableofcontents

\section{Introduction to ICMHD}\label{intro}

Consistent with the topic of this special volume, this paper will deal with the equations of
isentropic compressible magnetohydrodynamics (ICMHD) \cite{HKM1,Biskamp,PF}. Not only are 
these sufficiently general to encompass most of the interest in this area but they also offer 
a significant challenge even without dissipation. The ICMHD equations determine the 
dynamics of a conducting fluid flow with magnetic field $\bB$ satisfying $\hbox{div}\,\bB = 0$ 
and current $\curr = \hbox{curl}\,\bB$. The equations for the fluid velocity field $\bu$, mass 
density $\rho$, specific entropy $\sigma$, pressure $p(\rho,\sigma)$ and divergenceless magnetic 
field $\bB$ are, respectively, the motion equation, Faraday's Law of frozen-in magnetic flux,
isentropy along flow lines and the continuity equation\,:
\bel{mhd1a}
\rho \frac{D\bu}{Dt} = - \nabla p + \curr \times \bB\,,\hspace{2cm}
\frac{\partial\bB}{\partial t} = \hbox{curl}\,(\bu\times\bB)\,,
\ee
\bel{mhd1b}
\frac{D\sigma}{Dt} = 0\,,\hspace{2cm}
\frac{D\rho^{-1}}{Dt~~~} = \rho^{-1}{\rm div}\,\bu\,,
\ee
with the Lagrangian time derivative defined as
\bel{matderiv}
\frac{D~}{Dt} = \frac{\partial~}{\partial t} + \bu\cdot\nabla \,.
\ee 
The equation of state for specific internal energy $e(\rho,\sigma)$ and the
Thermodynamic First Law yield
\bel{1stLaw}
\frac{De}{Dt} 
= - \, p \frac{D\rho^{-1}}{Dt} + T \frac{D\sigma}{Dt}
= - \, \frac{p}{\rho}\,{\rm div}\,\bu
\,,
\ee
so the specific internal energy changes only because of mechanical work.  
Combining Faraday's Law with the continuity equation yields 
\bel{vect-fld}
\frac{D\bB_\rho}{Dt} -  \bB_\rho\cdot\nabla\bu = 0
\quad\hbox{with}\quad 
\bB_\rho:=\rho^{-1}\bB\,.
\ee
This is the condition for the vector field $\bB_\rho\cdot\nabla$ to be frozen into the flow,
i.e., 
\bel{frozen-vect-fld}
\frac{D}{Dt}\Big(\bB_\rho\cdot\frac{\partial}{\partial \bx}\Big) = 0
\,,\quad\hbox{along}\quad 
\frac{D\bx}{Dt}:=\bu\,.
\ee
We denote
\bel{sdef}
\varpi := p + \shalf B^{2}\,,
\ee
and set
\bel{sdef}
\frac{d}{ds}:=\bB_\rho\cdot\nabla\,,
\ee
with $\bB_{\rho}$ defined in (\ref{vect-fld}). In this notation, the ICMHD equations 
(\ref{mhd1a}) and (\ref{mhd1b}) transform using standard vector identities into
\bel{ICMHD1}
\frac{D\bu}{Dt} = \,\frac{d\bB}{ds} - \rho^{-1}\nabla\varpi =:\bF \,,
\hspace{1.5cm}
\frac{D\bB_\rho}{Dt}  = \frac{d\bu}{ds} \,,
\ee
\bel{ICMHD2}
\frac{D\sigma}{Dt} = 0
\hspace{3cm}
\frac{D\rho^{-1}}{Dt~~~}= \rho^{-1}{\rm div}\,\bu\,.
\ee
$D/Dt$ and $d/ds$ are defined in (\ref{matderiv}) and (\ref{sdef}) respectively. The first 
equation in (\ref{ICMHD1}) expresses the motion equation in terms of the derivative along 
field lines, while the second expresses Faraday's Law (\ref{vect-fld}) in terms of the 
derivatives $D/Dt$ and $d/ds$. The invariance condition (\ref{frozen-vect-fld}) for vector 
field $d/ds=\bB_\rho\cdot\nabla$ along Lagrangian field lines for ICMHD implies equality 
of the following cross derivatives,
\bel{Ertel}
\frac{D}{Dt}\frac{d}{ds} = \frac{d}{ds}\frac{D}{Dt}\,.
\ee
Of  course, the invariant vector field $d/ds=\bB_\rho\cdot\nabla$ may be applied to any fluid
quantity. For example, applying $d/ds$ to the equations in (\ref{ICMHD1}) and using the equality 
of cross derivatives in $d/ds$ and $D/Dt$ yields the following exact nonlinear wave equations 
for ICMHD
\begin{eqnarray}
\label{B-wave}
\frac{D^2(\rho^{-1}\bB)}{Dt^2} -\,\frac{d^2\bB}{ds^2}
&=&
-\,\frac{d}{ds}\Big( \rho^{-1}\nabla\varpi\Big)\,,\\
\frac{D^2\bu}{Dt^2} -\,\frac{d^2\bu}{ds^2}
&=& -\,\frac{D}{Dt}\Big( \rho^{-1}\nabla\varpi\Big) \,.
\label{u-wave}
\end{eqnarray}
When linearized, these equations yield Alfv\'en-sound waves. The present work emphasizes the
alignment dynamics that is inherent in these equations, rather than their wave properties.

\subsection{Slow plus fast decomposition and nonlinear waves} 

It seems natural to define slow and fast aspects of the ICMHD solutions in the sense of the
Lagrangian time derivative. In particular, there are two relations for slow variables, 
\bel{slow-var}
\frac{D\sigma}{Dt} = 0\,,\hspace{1cm}\hbox{and}\hspace{1cm}
\frac{D\beta}{Dt} = 0\,,
\ee
where $\beta$ is defined as
\bel{betadef}
\beta := \bB_{\rho}\cdot\nabla \sigma \,.
\ee 
The latter follows from equality of cross derivatives, by computing
\bel{Ertel-MHD}
\frac{D\beta}{Dt} := \frac{D}{Dt}\,(\bB_\rho\cdot\nabla \sigma) =
\frac{D}{Dt}\,\frac{d\sigma}{ds} = \frac{d}{ds} \, \frac{D\sigma}{Dt} = 0\,.
\ee
Thus, the ICMHD equations preserve the projection of $\bB_\rho$ on
$\nabla\sigma$ along {\em Lagrangian flow lines}, that is, along the path $\bx(t)$ of a
Lagrangian fluid particle determined from $d\bx/dt=\bu(\bx,\,t)$. Equivalently, one finds the
Eulerian  conservation law,
\bel{beta-CL}
\frac{\partial (\rho \beta)}{\partial t} + {\rm div}(\rho \beta \bu) = 0
\,,\quad\hbox{with}\quad
\rho \beta = \bB\cdot\nabla\sigma = {\rm div}(\sigma\bB)\,,
\ee
upon combining (\ref{Ertel-MHD}) with the continuity equation for mass.
\par\medskip
Of course, the evolution of the component of $\bB_\rho$ {\em perpendicular} to $\nabla\sigma$
is not so simple! One may decompose $\bB_\rho$ into components parallel and perpendicular to
$\nabla\sigma$ as
\bel{BS}
|\nabla\sigma|^2\bB_\rho 
= \beta\nabla\sigma + \bgamma\times\nabla\sigma
= |\nabla\sigma|^2(\bB_\rho^{\|} + \bB_\rho^{\perp})
\quad\hbox{with}\quad
\bgamma\cdot\nabla\sigma = 0\,,
\ee
so that
\bel{beta-gamma}
\beta = \bB_\rho\cdot\nabla\sigma 
\quad\hbox{and}\quad
\bgamma = \bB_\rho\times\nabla\sigma\,.
\ee
One then computes the auxiliary equations,
\bel{sig-dyn}
\frac{D}{Dt} \nabla\sigma
= - \, (\nabla\bu)^T\cdot\nabla\sigma\,,
\quad\hbox{that is,}\quad
\frac{D\sigma_{,i}}{Dt} = -\,\sigma_{,j}u^j_{,i}
\ee
and 
\bel{sig2dyn}
\frac{D}{Dt} |\nabla\sigma|^2 = 
- \nabla\sigma\cdot S \cdot \nabla\sigma
\ee
with the fluid strain-rate tensor defined as $S = \shalf(\nabla\bu + (\nabla\bu)^T)$.
Consequently, the evolution equation for the component of $\bB_\rho$ {\em
perpendicular} to $\nabla\sigma$ is determined from
\bel{gamma-dot}
\frac{D\bgamma}{Dt} =
\frac{D}{Dt} (\bB_\rho\times\nabla\sigma) =
\frac{d\bu}{ds} \times \nabla\sigma -
\bB_\rho \times (\nabla\bu^T\cdot\nabla\sigma)
\ee
which is clearly neither fast nor slow. 

\subsection{Lagrangian dynamics of specific volume} 

Two other likely candidates for fast variables are the specific volume
$\rho^{-1}$ and the velocity divergence ${\rm div}\,\bu$, which satisfy  
\bel{fast-var1}
\frac{D\rho^{-1}}{Dt~~~} = \rho^{-1}{\rm div}\,\bu\,,
\ee
\bel{fast-var2}
\frac{D}{Dt}{\rm div}\bu  = {\rm div}\bF - |\nabla\bu|^2 \,.
\ee
The second of these equations is found using the identity
\bel{fast-var3}
\frac{D}{Dt}{\rm div}\bu  = {\rm div}\Big(\frac{D\bu}{Dt}\Big) - |\nabla\bu|^2
\ee
where $|\nabla\bu|^2 \equiv u^i_{,j}u^j_{,i\,}$. As a consequence of (\ref{fast-var1}) 
and (\ref{fast-var2}), the specific volume $\rho^{-1}$ satisfies the ``Lagrangian 
oscillation equation,''
\bel{rho-osc}
\frac{D^2\rho^{-1}}{Dt\,^2~~} = \rho^{-1}
\left(({\rm div}\bu)^{2} - |\nabla\bu|^{2} + {\rm div}\Big(\frac{d\bB}{ds} -
\rho^{-1}\nabla\varpi\Big)\right)
\ee
upon substituting the definition of force $\bF:=\frac{d\bB}{ds} - \rho^{-1}\nabla\varpi$.
Thus, the divergence of the force combines with compressibility of the flow to drive either
Lagrangian oscillations or exponential variations of the specific volume, depending on the
sign of the term on the right hand side of equation (\ref{rho-osc}). Of course, the sign of
this term could be varying rapidly in some physical applications, and observations of such
behavior would have corresponding implications for the dynamics of specific volume. In any
case, the form of equation (\ref{rho-osc}) is universal and one may conclude that the
divergence of the total force drives the second Lagrangian time derivative of the specific
volume. 
\par\medskip
In what follows, we shall seek equations for alignment dynamics in MHD which have the
same universal property as the Lagrangian oscillation equation (\ref{rho-osc}). In
particular, we shall find that {\em gradients} of the total force (rather than its
divergence) drive the Lagrangian dynamics of alignment in MHD of the frozen-in magnetic
vector field $\bB_\rho$ relative to its projection $\bB_\rho\cdot\nabla
\bu$ onto the shear tensor $\nabla \bu$ of the fluid velocity. 

\subsection{Stretching and alignment}\label{stretch}

The magnetic flux equation
\bel{flux-eqn}
\frac{D\bB_\rho}{Dt}  = \frac{d\bu}{ds} := \bB_\rho\cdot\nabla \bu
\ee
implies that the contravariant vector $\bB_\rho$ undergoes stretching to the extent it aligns
with the shear $\nabla\bu$. Likewise, this alignment evolves according to
\bel{align-eqn}
\frac{D^2\bB_\rho}{Dt^2}  
= \frac{D}{Dt}\frac{d\bu}{ds}
= \frac{d}{ds}\frac{D\bu}{Dt}
= \frac{d}{ds}\bF
= \frac{d}{ds}\Big(\,\frac{d\bB}{ds} - \rho^{-1}\nabla\varpi\Big)\,,
\ee
which recovers the nonlinear wave equation (\ref{B-wave}) above. 
\par\medskip
We shall examine this equation from an alignment viewpoint, rather than as a wave propagation 
phenomenon. We begin by decomposing the vector $\bB_\rho\cdot\nabla \bu$ into its components 
parallel and perpendicular to $\bhB_\rho:=\bB_\rho/|\bB_\rho|$
\bel{par-perp-decomp}
\bB_\rho\cdot\nabla \bu =
\alpha\, \bhB_\rho + \bchi\times \bhB_\rho\,,
\ee
where $\alpha$ and $\bchi$ are defined by
\bel{acdef}
\alpha = \bhB_{\rho}\cdot(\bhB_{\rho}\cdot\nabla\bu)\,,
\hspace{1cm}\hbox{and}\hspace{1cm}
\bchi = \bhB_{\rho}\times(\bhB_{\rho}\cdot\nabla\bu) \,.
\ee
This decomposition is explained more fully in \S\ref{gendecom}. In fact, $\alpha$ is the 
Lagrangian amplification rate of the magnitude 
$|\bB_\rho|$ and $\bchi$ is the Lagrangian frequency of rotation of the unit vector 
$\bhB_\rho$ under the forcing by shear in the flux conservation equation (\ref{flux-eqn}),
\bel{alpha-chi-defs}
\frac{D |\bB_\rho|}{Dt} = \alpha\, |\bB_\rho|
\hspace{2cm}\hbox{and}\hspace{2cm}
\frac{D \bhB_\rho}{Dt} = \bchi \times \bhB_{\rho}\,.
\ee
Note that no confusion should arise between this $\alpha$ and the one appearing in the
$\alpha-$dynamo equations!

\subsection{A mathematical framework for magnetic fluid alignment dynamics}\label{framework}

The quantity $\alpha$ defined in equation (\ref{acdef}) is the {\em alignment} of $\bB_\rho$ with 
$\nabla\bu$, which determines the rates of change of magnitude $|\bB_\rho|$ whereas $\bchi$ is 
the {\em misalignment}, which determines the frequency of rotation of the direction $\bhB_{\rho}$. 
The question asked  in this situation is,
``How long will $\bB_\rho$ remain aligned with $\nabla\bu$, so it can continue to be
stretched under the fluid shear?'' Answering this question requires an analysis of
$D\alpha/Dt$ and $D\bchi/Dt$ using the alignment dynamics (\ref{align-eqn}). 
\par\medskip
In principle, this analysis could be performed by direct computation and algebraic
manipulation. However, the Lagrangian evolution equations for $\alpha$ and $\bchi$ happen to
fit perfectly into a mathematical framework that was especially designed for analyzing
orientation dynamics and for systematically interpreting the results. The key for recognizing
this framework is to notice that the decomposition of a vector into its components parallel
and perpendicular to another vector defines a type of product, or multiplication, that was
first discovered by Hamilton \cite{Ha1843}. This product reveals itself when we write the
parallel-perpendicular vector decomposition equation (\ref{par-perp-decomp}) as though it
were the pure vector components of a equation involving the four-component scalar-vector
object (tetrad) $[\alpha,\,\bchi]$ in the form,
\bel{par-perp-decomp-quat}
[0,\bB_\rho\cdot\nabla \bu] =
[0,\,\alpha\, \bhB_\rho + \bchi\times \bhB_\rho] =
[\alpha,\,\bchi]\circledast [0,\, \bhB_\rho]
\quad\hbox{with}\quad
\bchi\cdot\bhB_\rho=0\,.
\ee
This organization of the decomposition equation (\ref{par-perp-decomp}) 
summons the $\cast$ product defined by 
\bel{quat-prod}
[p_1,\,\bq_1]\cast[p_2,\,\bq_2] 
= [p_{1}p_{2} - \bq_{1}\cdot\bq_{2},\,p_{1}\bq_{2} + \bq_{1}p_{2}
+ 
\bq_{1}\times\bq_{2}]\,,
\ee
which is the multiplication rule that Hamilton invented for the field of quaternions 
\cite{Ha1843}. The origin of this rule and its connection to the definition of a 
quaternion is given in \S\ref{back}. Of course, quaternions have everything to do 
with orientation \cite{Ha2006,Ku1999}. The remainder of this paper sets up the quaternionic
framework for alignment dynamics of a general fluid flow and applies it to the ICMHD
equations.
\par\medskip
The plan of this paper is as follows: \S\ref{lagev} summarizes the results of \cite{GH06a} and 
shows how the evolution of  the ortho-normal frame can be calculated in general terms.
\S\ref{App-ICMHD} then discusses the ICMHD equations (\ref{ICMHD1}) and (\ref{ICMHD2}) in these 
terms, while \S\ref{align} examines  alignment and growth properties to begin answering the 
question raised above for magnetic dynamos using these results. Finally, \S\ref{conclus} 
discusses the potential for other applications of this general method in compressible MHD 
turbulence.

\section{Quaternions and Lagrangian alignment dynamics}\label{lagev}

\subsection{Background for quaternions}\label{back}

The resurgence of practical interest in quaternions during the last two decades has been 
stimulated by progress in the computer animation and inertial navigation industries because
of the ease with which quaternions handle moving objects undergoing three-axis rotations 
\cite{Ha2006,Ku1999}. The tracking of aircraft and satellites and the animation of tumbling 
objects in computer graphics are obvious examples. Quaternion methods have also been recently
been  applied to the three-dimensional Euler equations for incompressible fluid motion
\cite{GHKR,Gibbon02} and to passive tracer  particles transported by an underlying Lagrangian 
flow field; see \cite{GKB,LPVCAB2001,VLPCBA2002}  and references in \cite{GH06a}.  
The final result of these endeavors is that equations of motion can be derived for an 
ortho-normal frame. This `quaternion frame' follows the evolution of particles in a
Lagrangian  flow whose evolution derives from the Eulerian equations of motion.
\par\medskip
Three-axis rotations lie at the heart of the definition of a quaternion \cite{Ta1890}. In 
terms of any scalar $p$ and any 3-vector $\bq$, the quaternion $\bfq = 
[p,\,\bq]$ is defined (using Gothic fonts to denote quaternions) as
\bel{quatdef1}
\bfq = [p,\,\bq] = pI - \sum _{i=1}^{3}q_{i}\sigma_{i}\,,
\ee
where $\{\sigma_{1},\,\sigma_{2},\,\sigma_{3}\}$ are the Pauli spin-matrices 
and $I$ is the unit matrix. The relations between the Pauli matrices
$\sigma_{i}\sigma_{j} =  -\delta_{ij}I-\epsilon_{ijk}\sigma_{k}$ then give a 
non-commutative multiplication rule
\bel{quatdef2}
\bfq_{1}\cast\bfq_{2} = [p_{1}p_{2} - \bq_{1}\cdot\bq_{2},\,p_{1}\bq_{2} + \bq_{1}p_{2} + 
\bq_{1}\times\bq_{2}]\,.
\ee
It is easily demonstrated that quaternions are associative. In fact the individual 
elements of a unit quaternion provide the Cayley-Klein parameters of a rotation. This 
representation is a standard alternative to Euler angles in describing the orientation 
of rotating objects, as the books by Whittaker and Klein show \cite{Whitt1944,Klein04}.

\subsection{Quaternions and Lagrangian alignment dynamics in fluids}\label{gendecom}

A general quaternionic picture of the process of Lagrangian flow and acceleration in fluid
dynamics is explained in this  section by considering the abstract Lagrangian flow equation
\bel{w-dyn}
\frac{D\bw}{Dt} = \ba(\bx,\,t)\,,
\ee
whose Lagrangian acceleration equation is given in general by
\bel{a-dyn}
\frac{D^2\bw}{Dt^2} = \frac{D\ba}{Dt} = \bdb(\bx,\,t)\,.
\ee
These are the rates of change of these vectors following the characteristics of the velocity
generating  the path $\bx(t)$ of a Lagrangian fluid particle determined from
$d\bx/dt=\bu(\bx,\,t)$. 

\remfigure{
\par\vspace{-3mm}\noindent
\bc
\begin{minipage}[c]{.75\textwidth}
\begin{pspicture}
\psframe(0,0)(5,5)
\thicklines
\qbezier(0,1)(4,2.5)(0,4)
\thinlines
\put(2.01,2.43){\makebox(0,0)[b]{$\bullet$}}
\put(1.4,2.5){\makebox(0,0)[b]{\scriptsize$(\bx_{1},t_{1})$}}
\thinlines
\put(2,2.5){\vector(0,1){1}}
\put(2,3.7){\makebox(0,0)[b]{$\bhw$}}
\put(2,2.5){\vector(-2,-1){1}}
\put(.7,1.8){\makebox(0,0)[b]{$\bhchi_{a}$}}
\put(2,2.5){\vector(1,0){1}}
\put(3.8,2.4){\makebox(0,0)[b]{$\bhw\times\bhchi_{a}$}}
\thicklines
\qbezier(7,1)(6,2.5)(8,4)
\thinlines
\put(6.77,2.45){\makebox(0,0)[b]{$\bullet$}}
\put(6.2,2.5){\makebox(0,0)[b]{\scriptsize$(\bx_{2},t_{2})$}}
\thicklines
\qbezier[50](2,2.5)(3,.5)(6.7,2.5)
\thinlines
\put(6.73,2.5){\vector(1,4){.3}}
\put(7,3.8){\makebox(0,0)[b]{$\bhw$}}
\put(6.7,2.5){\vector(4,-1){1}}
\put(8.4,2.8){\makebox(0,0)[b]{$\bhw\times\bhchi_{a}$}}
\put(6.7,2.5){\vector(4,1){1}}
\put(8.1,2.1){\makebox(0,0)[b]{$\bhchi_{a}$}}
\put(3,1.2){\vector(1,0){.6}}
\put(4,.7){\makebox(0,0)[b]{\small particle trajectory}}
\put(4.5,1.4){\vector(4,1){.6}}
\thinlines
\end{pspicture}
\end{minipage}
\ec
\bc
\vspace{-1mm}
\begin{minipage}[r]{\textwidth}
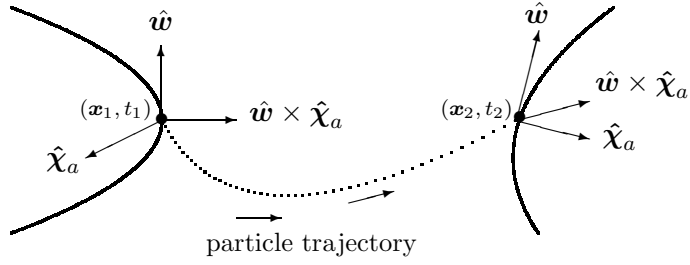
\textbf{Figure 1:} {\small The dotted line represents the path of Lagrangian
fluid particle $(\bullet)$ moving from $(\bx_{1},t_{1})$ to $(\bx_{2},t_{2})$. The solid
curves represent lines of constant  $\bw$ to which $\bhw$ is a unit tangent vector. The 
orientation of the quaternion-frame $(\bhw,\bhchi_{a},~\bhw\times\bhchi_{a})$ is  shown at
the two space-time points; note that this  is not the Frenet-frame corresponding to the
particle path but to lines of constant $\bw$.}
\end{minipage}
\ec
}
\par\medskip\noindent
Given the Lagrangian equation (\ref{w-dyn}) one defines the scalar $\alpha_{a}$ and the
3-vector $\bchi_{a}$  as
\bel{la2a}
\alpha_{a} = w^{-1}(\bhw\cdot\ba)\,,\hspace{2cm}\bchi_{a} = w^{-1}(\bhw\times\ba)\,,
\ee
in which $\bw=w\bhw$ with $w = |\bw|$. As observed in (\ref{par-perp-decomp}), the 3-vector $\ba$ 
is decomposed into parts that are parallel and perpendicular to $\bw$ as
\bel{decom1}
\ba = \alpha_{a}\bw + \bchi_{a}\times\bw 
= [\alpha_{a},\,\bchi_{a}]\cast[0,\,\bw]\,,
\ee
and thus the quaternionic product (\ref{decom1}) is summoned in a natural manner. By
definition, the growth rate $\alpha_{a}$  of the scalar magnitude $w = |\bw|$ obeys 
\bel{la3}
\frac{Dw}{Dt} = \alpha_{a}w\,,
\ee
while the unit tangent vector $\bhw = \bw w^{-1}$ satisfies 
\bel{la5}
\frac{D\bhw}{Dt} = \bchi_{a}\times \bhw\,.
\ee
Now define two quaternions 
\bel{ls6}
\bfq_{a} 
= [\alpha_{a},\,\bchi_{a}]
\quad\hbox{and}\quad
\bfq_{b} 
= [\alpha_{b},\,\bchi_{b}]\,,
\ee
where $\alpha_{b},~\bchi_{b}$ are defined as in (\ref{la2a}) for
$\alpha_{a},~\bchi_{a}$  with $\ba$ replaced by $\bdb$. Let $\bfw =
[0,\,\bw]$ be the pure quaternion satisfying  the Lagrangian evolution
equation (\ref{w-dyn}) with $\bfq_{a}$ defined in (\ref{ls6}).  Then 
(\ref{w-dyn}) can automatically be re-written equivalently in the 
quaternion form
\bel{lem1}
\frac{D\bfw}{Dt} 
= [0,\,\ba] 
= [0,\, \alpha_{a}\bw + \bchi_{a}\times\bw ] 
= \bfq_{a}\cast\bfw\,.
\ee
Moreover, if $\ba$ is  Lagrangian-differentiable as in
(\ref{a-dyn}) then it  is clear that a similar decomposition for $\bdb$ as
that for $\ba$ in (\ref{decom1}) gives 
\bel{la9}
\frac{D^{2}\bfw}{Dt^{2}} 
= [0,\,\bdb] 
= [0,\, \alpha_{b}\bw + \bchi_{b}\times\bw ] 
= \bfq_{b}\cast\bfw \,.
\ee
Using the associativity property, compatibility of (\ref{la9}) 
and (\ref{lem1}) implies that
\bel{la10}
\left(\frac{D\bfq_{a}}{Dt} + \bfq_{a}\cast\bfq_{a} 
-\bfq_{b}\right)\cast\bfw = 0\,,
\ee
which establishes a \textit{Riccati relation} between the quaternions $\bfq_{a}$ and
$\bfq_{b}$
\bel{Ric1}
\frac{D\bfq_{a}}{Dt} + \bfq_{a}\cast\bfq_{a} = \bfq_{b}\,.
\ee
From equation (\ref{Ric1}) there follows the main result of the paper:
\begin{theorem}\label{abthm}
The ortho-normal quaternion-frame $F=(\bhw,\,\bhchi_{a},\,\bhw\times\bhchi_{a})\in SO(3)$ 
has Lagrangian time derivatives expressed as
\beq{abframe3}
\frac{D\bhw}{Dt}&=& \bD_{a}\times\bhw\,,\\
\frac{D(\bhw\times\bhchi_{a})}{Dt} &=& \bD_{a}\times(\bhw\times\bhchi_{a})\,,\label{abframe4}
\\
\frac{D\bhchi_{a}}{Dt} &=& \bD_{a}\times\bhchi_{a}\,,\label{abframe5}
\eeq
where the \textbf{Darboux vector} $\bD_{a}$ defined as
\bel{abframe6}
\bD_{a} = \frac{c_{b}}{\chi_{a}}\bhw + \bchi_{a}
\quad\hbox{with}\quad
c_{b} = \bchi_{b}\cdot(\bhw\times\bhchi_a)
\,,
\ee
is the angular frequency of rotation of the ortho-normal frame $F$.    
\end{theorem}
{\bfi Remarks.}
\begin{itemize}
\item The frame orientation is controlled by the Darboux vector $\bD_{a} = 
(c_b/\chi_a,\,0,\,\chi_a)$ which lies in the
$(\bhw,\,\bhchi_{a})$ plane and is so named for its similarity to the Darboux vector in the
Frenet-Serret equations for a space curve. Note that the vector $\bD_a$ depends on $\bchi_{b}$
but is independent of $\alpha_b$.
\item The frame dynamics equations (\ref{abframe3})-(\ref{abframe6}) may also be 
re-written in matrix form by defining the $3\times3$ skew-symmetric matrix $C_a\in so(3)$ with 
entries $[C_a]_{ij}=-\,\epsilon_{ijk}D_a^k$ as,
\bel{SO3-frame}
\frac{DF}{Dt} = C_aF
\quad\hbox{where}\quad
F = \left(
\begin{array}{c}
 \bhw    \\
  \bhw\times\bhchi_{a}   \\
\bhchi_{a}
\end{array}
\right)
\quad\hbox{and}\quad
C_a = \left(
\begin{array}{ccc}
0       & -\chi_a         & 0          \\
\chi_a       & 0          & -c_b/\chi_a \\
0       & c_b/\chi_a      & 0
\end{array}
\right)
\ee
\end{itemize}

\textbf{Proof\,:} Finding an expression for the Lagrangian time derivatives of the components
of  the frame  $(\bhw,\,\bhw\times\bhchi_{a},\,\bhchi_{a})^T$ requires the derivative of
$\bhchi_{a}$.  For this, one first recalls that the 3-vector $\bdb$ may be
expressed in this  ortho-normal frame as the linear combination
\bel{b1}
w^{-1}\bdb = \alpha_{b}\,\bhw + c_{b}\bhchi_{a} + d_{b}(\bhw\times\bhchi_{a})
\,,
\ee
where $c_{b}$ is defined in (\ref{abframe6}) and $d_{b} = -\,(\bhchi_{a}\cdot\bchi_{b})$.
The 3-vector product $\bchi_{b} = w^{-1}(\bhw\times\bdb)$ yields 
\bel{bchibdef}
\bchi_{b} = c_{b}(\bhw\times\bhchi_{a}) - d_{b}\bhchi_{a}\,.
\ee
To find the Lagrangian time derivative of $\bhchi_{a}$, we use the 3-vector part of the 
equation for the quaternion $\bfq_{a} = [\alpha_{a},\,\bchi_{a}]$ in Theorem \ref{abthm}
\bel{abframe1}
\frac{D\bchi_{a}}{Dt} = - 2\alpha_{a}\bchi_{a} + \bchi_{b}\,,
\hspace{1cm}
\Rightarrow
\hspace{1cm}
\frac{D\chi_{a}}{Dt} = -2\alpha_{a}\chi_{a} - d_{b}\,,
\ee
where $\chi_{a} = |\bchi_{a}|$. Using (\ref{bchibdef}) and (\ref{abframe1}) there follows
\bel{abframe2}
\frac{D\bhchi_{a}}{Dt} = c_{b}\chi_{a}^{-1}(\bhw\times\bhchi_{a})\,,
\hspace{2cm}
\frac{D(\bhw\times\bhchi_{a})}{Dt} = \chi_{a}\,\bhw - c_{b}\chi_{a}^{-1}\bhchi_{a}\,,
\ee
which gives equations (\ref{abframe3})-(\ref{abframe6}).\hspace{8cm}$\blacksquare$

\section{Application of the quaternionic alignment theorem to ICMHD}\label{App-ICMHD}

The way is now clear to apply Theorem \ref{abthm} of \S\ref{lagev} to the ICMHD equations
(\ref{ICMHD1}) and (\ref{ICMHD2}). First, we identify from (\ref{flux-eqn}) and (\ref{align-eqn}) 
\bel{more2}
\bw = \bB_{\rho}\,,\quad
\ba =\bB_{\rho}\cdot\nabla\bu \,,\quad
\bdb = \frac{D(\bB_{\rho}\cdot\nabla\bu)}{Dt} 
= \bB_{\rho}\cdot\nabla\left(\frac{D\bu}{Dt}\right)\,.
\ee
Although messy, this provides an explicit expression for $\bdb$ that depends 
upon the pressure $p$, $\bB_{\rho}$, $\curr$ and their derivatives, as determined from the
fluid evolution equations for $\bu$, $\rho$ and $\sigma$, 
\bel{more3}
\bdb 
=  \bB_{\rho}\cdot\nabla\left(-\rho^{-1}\nabla p + \curr\times\bB_{\rho}\right)\,.
\ee
The alignment parameters $\{\alpha,\,\bchi,\,\alpha_{b},\,\bchi_{b}\}$ are now identified as 
\bel{more4a}
\alpha = \bhB_{\rho}\cdot(\bhB_{\rho}\cdot\nabla\bu)
\,,\hspace{2cm}
\bchi = \bhB_{\rho}\times(\bhB_{\rho}\cdot\nabla\bu)\,,
\ee
\bel{more4b}
\alpha_{b} = |\bB_{\rho}|^{-1}(\bhB_{\rho}\cdot\bdb)
\,,\hspace{2cm}
\bchi_{b} =|\bB_{\rho}|^{-1}(\bhB_{\rho}\times\bdb)
\,.
\ee
These parameters appear in the vector decompositions, 
\bel{MHD-align1}
\ba = \bB_\rho\cdot\nabla \bu =
\alpha\, \bhB_\rho + \bchi\times \bhB_\rho\,,
\ee
and
\bel{MHD-align2}
\bdb = \bB_{\rho}\cdot\nabla\left(\frac{D\bu}{Dt}\right) 
= \alpha_b\, \bhB_\rho + \bchi_b\times \bhB_\rho\,.
\ee
The parameters $\alpha_b$ and $\bchi_b$ derive from $\bB_{\rho}$ and from $\bdb$ in
equation (\ref{more3}) at each time step. The vector $\bdb$ represents the coupling of the
kinematic flow variables $(\bB_{\rho},\nabla\bu)$ to the {\em gradients} of the magnetic and
thermodynamic forces. These identifications enter the two quaternions
$\bfq = [\alpha,\,\bchi]$ and
$\bfq_{b} = [\alpha_{b},\,\bchi_{b}]$ that satisfy the Riccati equation (\ref{Ric1}). The
results of Theorem \ref{abthm} for the Lagrangian time derivatives of the orthonormal frame 
$F_{mag} =(\bhB_\rho\,,\bhB_\rho\times\bhchi\,,\bhchi)^T\in SO(3)$ are then expressed as
\beq{more5a}
\frac{D\bhB_\rho}{Dt}&=& \bD\times\bhB_\rho\,,\\
\frac{D(\bhB_\rho\times\bhchi)}{Dt} &=& \bD\times(\bhB_\rho\times\bhchi)\,,\label{more5b}
\\
\frac{D\bhchi}{Dt} &=& \bD\times\bhchi\,,\label{more5c}
\eeq
where the Darboux angular velocity vector $\bD$ is defined as
\bel{more5d}
\bD = \frac{c_{b}}{\chi}\bhB_{\rho} + \bchi
\,,\quad\hbox{with}\quad
c_{b} = \bchi_{b}\cdot(\bhB_\rho\times\bhchi)
\,,
\ee
which depends explicitly on the ICMHD force gradients through $\bchi_{b}$ but is independent
of  $\alpha_{b}$.
\par\medskip
The frame dynamics equations (\ref{more5a})-(\ref{more5d}) for ICMHD may also be re-written in
matrix form using the $3\times3$ skew-symmetric matrix $C\in so(3)$ with entries
$C_{ij}=-\,\epsilon_{ijk}D^k$ as,
\bel{SO3-frame}
\frac{DF_{mag}}{Dt} = CF_{mag}
\quad\hbox{where}\quad
F_{mag} = \left(
\begin{array}{c}
 \bhB_\rho    \\
  \bhB_\rho\times\bhchi   \\
\bhchi
\end{array}
\right)
\quad\hbox{and}\quad
C = \left(
\begin{array}{ccc}
0       & -\chi         & 0          \\
\chi       & 0          & -c_b/\chi \\
0       & c_b/\chi      & 0
\end{array}
\right)
\ee
{\bfi Remarks.}
\begin{itemize}
\item
The skew-symmetric matrix  $C=\frac{DF_{mag}}{Dt}F_{mag}^{-1}$ expresses the Lagrangian
angular frequency of rotation of the magnetic orthonormal frame $F_{mag}$ in terms of the
magnitudes $\chi$ and $c_b$.
\item
At a given moment in time, the flow lines of $\bB_\rho$ may be constructed from its
characteristic equations $d\bx/ds=\bB_\rho(\bx)$. The Frenet-Serret (FS) equations describe 
how the orientation of an orthonormal frame changes along each flow line of $\bB_\rho$ as a
function of its shape parameters, curvature and torsion. Applying equality of cross
derivatives to equations (\ref{SO3-frame}) and (FS) yields a relation for the Lagrangian time
derivatives of the shape parameters of a given flow line of $\bB_\rho$. 
\end{itemize}

\remfigure{
Figure 1 now becomes 
\par\vspace{-3mm}\noindent
\bc
\begin{minipage}[c]{.75\textwidth}
\begin{pspicture}
\psframe(0,0)(5,5)
\thicklines
\qbezier(0,1)(4,2.5)(0,4)
\thinlines
\put(2.01,2.43){\makebox(0,0)[b]{$\bullet$}}
\put(1.4,2.5){\makebox(0,0)[b]{\scriptsize$(\bx_{1},t_{1})$}}
\thinlines
\put(2,2.5){\vector(0,1){1}}
\put(2,3.7){\makebox(0,0)[b]{\small$\bhB$}}
\put(2,2.5){\vector(-2,-1){1}}
\put(.7,1.8){\makebox(0,0)[b]{\small$\bhchi$}}
\put(2,2.5){\vector(1,0){1}}
\put(3.8,2.4){\makebox(0,0)[b]{\small$\bhB\times\bhchi$}}
\thicklines
\qbezier(7,1)(6,2.5)(8,4)
\thinlines
\put(6.77,2.45){\makebox(0,0)[b]{$\bullet$}}
\put(6.2,2.5){\makebox(0,0)[b]{\scriptsize$(\bx_{2},t_{2})$}}
\thicklines
\qbezier[50](2,2.5)(3,.5)(6.7,2.5)
\thinlines
\put(6.73,2.5){\vector(1,4){.3}}
\put(7,3.8){\makebox(0,0)[b]{\small$\bhB$}}
\put(6.7,2.5){\vector(4,-1){1}}
\put(8.4,2.8){\makebox(0,0)[b]{\small$\bhB\times\bhchi$}}
\put(6.7,2.5){\vector(4,1){1}}
\put(8.1,2.1){\makebox(0,0)[b]{\small$\bhchi$}}
\put(3,1.2){\vector(1,0){.6}}
\put(4,.7){\makebox(0,0)[b]{\small particle trajectory}}
\put(4.5,1.4){\vector(4,1){.6}}
\thinlines
\end{pspicture}
\end{minipage}
\ec
\bc
\vspace{-1mm}
\begin{minipage}[r]{\textwidth}
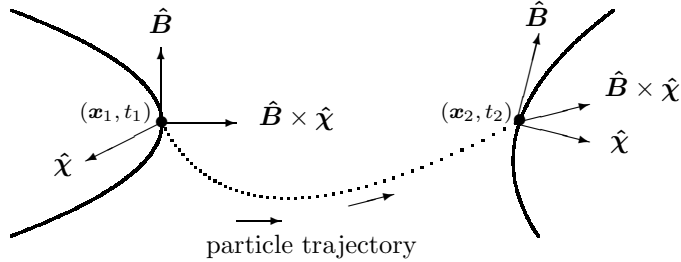
\textbf{Figure 2:} {\small  The solid curves represent magnetic field lines to which $\bhB$ is a 
unit tangent vector. The dotted line represents the path of a Lagrangian fluid particle
$(\bullet)$ moving from $(\bx_{1},t_{1})$ to $(\bx_{2},t_{2})$. The orientation of the
quaternion-frame $(\bhB,~\bhchi,~\bhB\times\bhchi)$ is shown at the two space-time points.}
\end{minipage}
\ec
}

\section{Alignment and growth properties in ICMHD}\label{align}

The growth rate $\alpha(\bx,\,t)$, defined in (\ref{more4a}), satisfies (see equations (\ref{la3}) 
and (\ref{la5}))
\bel{alig1}
\frac{D|\bB_{\rho}|}{Dt} = \alpha|\bB_{\rho}|\,.
\ee
Here $\alpha$ can take either sign and this is the key to how fast the magnitude 
$|\bB_{\rho}|$ increases (or decreases) at  each point in the flow.  In contrast, the 3-vector
$\bchi(\bx,\,t)$ is the key to the alignment  properties of the system, because it satisfies
\bel{alig2}
\frac{D\bhB_{\rho}}{Dt} = \bchi\times\bhB_{\rho}\,.
\ee
As such, it can be interpreted as the swing rate of the unit vector $\bhB_{\rho}$ about $\ba 
= \bB_{\rho}\cdot\nabla\bu$. Clearly, if $\bB_{\rho}$ is aligned with $\ba$ then $\bchi = 0$ 
and the quaternion $\bfq$ involves only the scalar $\alpha$. Violent corkscrew-like motions 
of the magnetic field lines would be consistent with significant values of $\bchi$ 
and such motions can therefore be regarded as a diagnostic for the misalignment of
$\bhB_{\rho}$  with $\ba$. One of the messages of this paper is that putting $\alpha$ and
$\bchi$ together  as the quaternion $\bfq = [\alpha,\,\bchi]$ is a natural way to approach
this problem  because the full quaternion $\bfq$ with $\bchi\ne 0$ is summoned whenever
vortex or magnetic field lines bend or tangle.  The Darboux vector $\bD$ is the angular
frequency of rotation of the orthonormal frame  $F_{mag} =
(\bhB_\rho\,,\bhB_\rho\times\bhchi\,,\bhchi)^T$,  but is itself controlled by the gradients
of forces in the expression (\ref{more3}) for $\bdb$.
\par\medskip
When written out in terms of $\alpha$ and $\bchi$, the Riccati relation (\ref{Ric1}) becomes
\bel{Ricc-vect}
\frac{D\alpha}{Dt} = \chi^{2}-\alpha^{2} + \alpha_{b}\,,
\qquad\frac{D\bchi}{Dt} = -2\alpha\bchi + \bchi_{b}\,.
\ee
Some years ago, these four equations were first expressed in this form  for the incompressible 
Euler equations without any recourse to quaternions \cite{GGH}. The quaternionic form first
appeared in \cite{Gibbon02} (see \cite{GHKR,GH06a} for a history). Equations (\ref{Ricc-vect}) 
appear to behave as Lagrangian ODEs driven by $\bfq_b = [\alpha_{b},\,\bchi_{b}]$.  If the latter 
terms remain roughly constant equations
(\ref{Ricc-vect}) can easily be shown to have two fixed points \cite{GGH}; one has a
negative value of $\alpha$ and the other has a positive value. The negative one is associated
with an unstable spiral and the positive one with a stable spiral in the phase plane for this
simplified ODE system. Equations (\ref{Ricc-vect}) with constant $\bfq =
[\alpha_{b},\,\bchi_{b}]$ have also recently been studied as a remarkably interesting
kinematic model for the creation of non-Gaussian statistics in hydrodynamic turbulence
\cite{LiMe2005}. This simple picture would seem not apply if the driving terms
$\alpha_{b},\,\bchi_{b}$ were to vary on the same time
scales as $\alpha,\,\bchi$, or faster. This breakdown in applicability of equations
(\ref{Ricc-vect}) with constant $\bfq = [\alpha_{b},\,\bchi_{b}]$ would be indicated if the
force gradients in (\ref{more3}) were observed to undergo rapid changes.

\section{Conclusion}\label{conclus}
The quaternionic approach to Lagrangian frame dynamics developed here for ICMHD applies
generally in fluid dynamics. Because the form of these equations is universal; that is,
independent of the specific choice of forces, one may expect them to have many other
applications. For example, in MHD one may use these equations to consider the effects on
frame dynamics of rotation, or the effects of various kinds of subgrid-scale models, simply
by identifying the corresponding expressions for the vectors $\ba$ and $\bdb$ in equations
(\ref{w-dyn}) and (\ref{a-dyn}). Subgrid-scale models of MHD turbulence may be a particularly
fruitful arena for these applications, especially for those models derived from Lagrangian
averaging, because Faraday's Law for such models is preserved for the averaged field
\cite{GrHoMiPo2006}. Depending as it does on Faraday's Law, the quaternionic method for MHD is
fundamental, but it is also mainly kinematic.  Thus its best role may be as a
means of developing diagnostics for determining the effects of total force gradients.
Therefore, additional applications of the quaternionic approach may also be foreseen, as
improved Lagrangian diagnostic methods are developed in the future. For example, future
diagnostics may be able to distinguish between effects described by equations
(\ref{Ricc-vect}) with fixed values of the quaternion
$\mathfrak{q}_b=[\alpha_{b},\,\bchi_{b}]$, versus its self-consistent exact
dynamics when the vector $\bdb$ is determined from the varying force gradients in equation
(\ref{more3}).

\par\vspace{4mm}\noindent
\textbf{Acknowledgements:} The work of DDH was partially supported  by the US Department 
of Energy, Office of Science, Applied Mathematical Research. 

\bibliographystyle{unsrt}


\end{document}